\documentclass[psfig,12pt,citesort]{article}  
\newcommand{\hoch}[1]{$\, ^{#1}$}

\def \ov {\over}

\def \p {\phi}  
  
\def \te {\tilde \epsilon}

\newcounter{subequation}[equation]

\newcommand{\be}{\begin{equation}}  
\newcommand{\ee}{\end{equation}}  
\newcommand{\eel}[1]{\label{#1}\end{equation}}  
\newcommand{\bea}{\begin{eqnarray}}  
\newcommand{\eea}{\end{eqnarray}}   
\newcommand{\eeal}[1]{\label{#1}\end{eqnarray}}  
\makeatletter  
\def\thesubequation{\theequation\@alph\c@subequation}  
\def\@subeqnnum{{\rm (\thesubequation)}}  
\def\slabel#1{\@bsphack\if@filesw {\let\thepage\relax  
   \xdef\@gtempa{\write\@auxout{\string  
      \newlabel{#1}{{\thesubequation}{\thepage}}}}}\@gtempa  
   \if@nobreak \ifvmode\nobreak\fi\fi\fi\@esphack}  
\def\subeqnarray{\stepcounter{equation}  
\let\@currentlabel=\theequation\global\c@subequation\@ne  
\global\@eqnswtrue \global\@eqcnt\z@\tabskip\@centering\let\\=\@subeqncr  
  
$$\halign to \displaywidth\bgroup\@eqnsel\hskip\@centering  
  $\displaystyle\tabskip\z@{##}$&\global\@eqcnt\@ne  
  \hskip 2\arraycolsep \hfil${##}$\hfil  
  &\global\@eqcnt\tw@ \hskip 2\arraycolsep  
  $\displaystyle\tabskip\z@{##}$\hfil  
   \tabskip\@centering&\llap{##}\tabskip\z@\cr}  
\def\endsubeqnarray{\@@subeqncr\egroup  
                     $$\global\@ignoretrue}  
\def\@subeqncr{{\ifnum0=`}\fi\@ifstar{\global\@eqpen\@M  
    \@ysubeqncr}{\global\@eqpen\interdisplaylinepenalty \@ysubeqncr}}  
\def\@ysubeqncr{\@ifnextchar [{\@xsubeqncr}{\@xsubeqncr[\z@]}}  
\def\@xsubeqncr[#1]{\ifnum0=`{\fi}\@@subeqncr  
   \noalign{\penalty\@eqpen\vskip\jot\vskip #1\relax}}  
\def\@@subeqncr{\let\@tempa\relax  
    \ifcase\@eqcnt \def\@tempa{& & &}\or \def\@tempa{& &}  
      \else \def\@tempa{&}\fi  
     \@tempa \if@eqnsw\@subeqnnum\refstepcounter{subequation}\fi  
     \global\@eqnswtrue\global\@eqcnt\z@\cr}  
\let\@ssubeqncr=\@subeqncr  
\@namedef{subeqnarray*}{\def\@subeqncr{\nonumber\@ssubeqncr}\subeqnarray}  
  
\@namedef{endsubeqnarray*}{\global\advance\c@equation\m@ne%  
                           \nonumber\endsubeqnarray}  
  
\makeatletter \@addtoreset{equation}{section} \makeatother  
\renewcommand{\theequation}{\thesection.\arabic{equation}}  
\catcode`\@=11  
  
\newcount\hour  
\newcount\minute  
\newtoks\amorpm \hour=\time\divide\hour by 60\minute  
=\time{\multiply\hour by 60 \global\advance\minute by-\hour}  
\edef\standardtime{{\ifnum\hour<12 \global\amorpm={am}%  
        \else\global\amorpm={pm}\advance\hour by-12 \fi  
        \ifnum\hour=0 \hour=12 \fi  
        \number\hour:\ifnum\minute<10  
        0\fi\number\minute\the\amorpm}}  
\edef\militarytime{\number\hour:\ifnum\minute<10 0\fi\number\minute}  
  
\def\draftlabel#1{{\@bsphack\if@filesw {\let\thepage\relax  
   \xdef\@gtempa{\write\@auxout{\string  
      \newlabel{#1}{{\@currentlabel}{\thepage}}}}}\@gtempa  
   \if@nobreak \ifvmode\nobreak\fi\fi\fi\@esphack}  
        \gdef\@eqnlabel{#1}}  
\def\@eqnlabel{}  
\def\@vacuum{}  
\def\marginnote#1{}  
\def\draftmarginnote#1{\marginpar{\raggedright\scriptsize\tt#1}}  
\def\draft{  
        \pagestyle{plain}  
        \overfullrule=2pt  
        \oddsidemargin -.5truein  
        \def\@oddhead{\sl \phantom{\today\quad\militarytime} \hfil  
        \smash{\Large\sl DRAFT} \hfil \today\quad\militarytime}  
        \let\@evenhead\@oddhead  
        \let\label=\draftlabel  
        \let\marginnote=\draftmarginnote  
        \def\ps@empty{\let\@mkboth\@gobbletwo  
        \def\@oddfoot{\hfil \smash{\Large\sl DRAFT} \hfil}  
        \let\@evenfoot\@oddhead}  
  
\def\@eqnnum{(\theequation)\rlap{\kern\marginparsep\tt\@eqnlabel}%  
        \global\let\@eqnlabel\@vacuum}  }

\renewcommand{\theequation}{\thesection.\arabic{equation}}  
\renewcommand{\thefootnote}{\fnsymbol{footnote}}

\def\appendix#1{  
  \addtocounter{section}{-3}  
  \setcounter{equation}{0}  
  \renewcommand{\thesection}{\Alph{section}}  
  \section*{Appendix \thesection\protect\indent \parbox[t]{11.15cm}  
  {#1} }  
  \addcontentsline{toc}{section}{Appendix \thesection\ \ \ #1}  
  }  
\def \ov {\over}

\def \p {\phi}  
  
\def \te {\tilde \epsilon}

\def\o{\omega}

\def\pd{\partial}

\def\m{\mu}  
\def\n{\nu}  
\def\a{\alpha}  
\def\b{\beta}  
\def\g{\gamma}

\def\r{\rho}

\def\s{\sigma}

\def\te{\theta}

\def\p{\phi}

\def\be{\begin{equation}}  
\def\ee{\end{equation}}

\def \g {\gamma}

\def \m {\mu}  
\def \n {\nu}

\def\te{\theta}  
\def\g{\gamma}

\date{}  
\begin{document}

%%%% Title page  
\begin{flushright}  
\hfill{\bf hep-th/0306107}\\  
  
\hfill{MCTP-03-28}\\  
\hfill{PUPT-2088}  
  
\end{flushright}  
  
\begin{center}

\vspace{2cm}  
  
{\Large {\bf Hadronic Density of States from String Theory}}  
  
\vspace{30pt}  
{\large Leopoldo A. Pando Zayas\hoch{1} and  
Diana Vaman\hoch{2}}  
  
\vspace{14pt}  
\hoch{1} {\it Michigan Center for Theoretical Physics,  
University of Michigan\\  
Ann Arbor, MI 48109--1120}\\  
{\tt lpandoz@umich.edu}

\vspace{14pt}  
\hoch{2} {\it Department of Physics,  
Princeton University\\  
Princeton, NJ 08544}\\  
{\tt dvaman@feynman.princeton.edu}  
  
\vspace{3.5cm}  
\underline{ABSTRACT}  
\end{center}
  
Exactly soluble string theories describing a particular hadronic sector
of certain confining gauge theories have been obtained recently as
Penrose-G\"uven limits of the dual supergravity backgrounds.  
The effect of taking the Penrose-G\"uven limit on the gravity side
translates, in the gauge theory side, into an effective 
truncation to hadrons of large U(1) charge
(annulons).
We present
an exact calculation of the finite temperature partition function for
the hadronic states corresponding to a Penrose-G\"uven limit 
of the
Maldacena-N\`u\~nez embedding
of ${\cal N}=1$ SYM into string theory. It is established that the
theory exhibits a Hagedorn density of states.

Motivated by this exact calculation we propose a semiclassical
string approximation to the finite temperature partition function for
confining gauge theories admitting a supergravity dual, by performing an 
expansion around classical solutions characterized by temporal windings. 
This semiclassical approximation reveals a hadronic energy density of
states of Hagedorn type, with the coefficient determined by the gauge
theory string tension as expected for confining theories. We argue
that our proposal captures primarily information about
states of pure ${\cal N}=1$ SYM, given that this semiclassical approximation 
does not entail a projection onto states of large U(1) charge.  
  
\newpage   
  
%\draft  

\setcounter{page}{1} \renewcommand{\thefootnote}{\arabic{footnote}}  
\setcounter{footnote}{0}  
  
\def \N{{\cal N}} \def \ov {\over}  
  
\setcounter{page}{1} \renewcommand{\thefootnote}{\arabic{footnote}}  
\setcounter{footnote}{0}  
  
\def \N{{\cal N}} \def \ov {\over}

%%%%%%%%%%%%%%%%%%%%%%%%%%%%%%%%%%%%%%%%%%%%%%%%%%%%%%%%%%%%%%%%%%%%%  
\section{Introduction }  
%%%%%%%%%%%%%%%%%%%%%%%%%%%%%%%%%%%%%%%%%%%%%%%%%%%%%%%%%%%%%%%%%%%

$${}$$
   
The duality between ${\cal N}=4$ Supersymmetric Yang-Mills (SYM)
theory and  
string theory in $AdS_5\times S^5$ \cite{ads} has given a concrete playground  
where ideas about the gauge/gravity correspondence have been widely  
tested. In a series of papers, some of them predating the AdS/CFT proposal, 
Klebanov  
and collaborators \cite{igor} established a direct relation  
between the entropy  of ${\cal N}=4$ SYM and the Bekenstein Hawking  
entropy of  a stack of near extremal D3-branes.   
  
Attempts at generalizing the ideas of \cite{igor} to  cases of  
supergravity backgrounds dual to confining gauge theories  
\cite{bh1,bh2} have encountered multiple obstacles. In particular,  
nonextremal generalizations of such supergravity backgrounds are  
conjectured to exists only for high enough temperatures after chiral  
symmetry is restored and the theory has settled into the  
deconfined phase. It is fair to say that the question of understanding the 
hadronic density  
of states has so far alluded a supergravity approach. An intuitive  
explanation for the failure of supergravity  to capture the  
density of states of confining gauge theories can be given in terms of  
the string partition function. Identifying the string partition
function with the gauge theory partition function, as instructed by the
gauge/gravity correspondence, the standard genus expansion in string  
theory predicts the following form of the partition function:   
$$  
Z_{string}=N^2\,Z_0+N^0\,Z_1+{1\over N^2}Z_2+\ldots,  
$$  
where we identify $g_s\equiv N^{-1}$. 
This implies that to understand  the confined phase (the $N^0$ term) one must 
consider  
strings with torus topology. From this point of view the deconfined 
quark-gluon  
contribution (the $N^2$ term)   
has been successfully understood at the supergravity level \cite{igor,bh1,bh2} via the  
Bekenstein Hawking entropy. In this paper we attempt to study the  
$Z_1$ term in the above expansion.   
  
In principle, determining $Z_1$ requires knowledge of the full string  
spectrum. Although the full super string theory in AdS-like backgrounds  remains a elusive goal,  
recently progress has been made by considering certain limits. In   
\cite{bmn} a dictionary was established between certain large R-charge  
operators in ${\cal N}=4$ SYM and string theory in the Penrose limit  
of $AdS_5\times S^5$. In a somewhat generalizing proposal \cite{gkp}, it
was argued that by way of studying classical configurations of the string  
sigma model in supergravity backgrounds one can obtain information  
about specific sectors of the spectrum of strings in such  
backgrounds. In particular, \cite{gkp} provided a   
semiclassical string derivation of the anomalous dimension of twist-two  
operators in ${\cal N}=4$ SYM. Interestingly, this relation holds for  
the corresponding twist-two operators in QCD. Other remarkable
results, like the presence of hard amplitudes for   
strings, have been obtained in a conceptually similar  
line of attack which approximates string theory processes in AdS-like  
backgrounds by a convolution of wave functions in the AdS-like  
background and standard string theory amplitudes \cite{ps}. This  
climate encourages us to looked for a semiclassical alternative to the  
computation of the finite temperature partition function.   
  
The structure of $Z_1$ above turns out to be intimately related to  
the nature of the Hagedorn density of states in the gauge/gravity  
correspondence. Ever since in the 60's the analysis of experimental data from  
hadron scattering lead Hagedorn \cite{hagedorn} to introduce the   
asymptotic  
  bootstrap condition  
(now known as a Hagedorn density of states), the nature of this  
distribution has been a source of interest. In the 80's, explicit  
computations of the partition functions of all consistent string  
theories, showed the universality of the Hagedorn density of states in  
string theory. More recently, based on the solubility of string theory   
on a curved plane wave background with Ramond Ramond flux  
\cite{exactstring}, a Hagedorn temperature has been  
established in a new exactly solvable string theory on a curved  
background with Ramond Ramond flux \cite{pv} (see also \cite{therest,gost,nothing}).  
The universal appearance   
of a Hagedorn density of states in string theories adds to the  
fascination of the subject. Several authors have studied this very  
tantalizing similarity between gauge and string theories in the  
AdS/CFT context \cite{rabinovici}.   
  
Our goal in this paper is twofold. First, we compute exactly the  
thermal partition function of a string theory description of certain  
hadronic states. By analyzing the structure of the result we are lead  
to a proposal which constitutes the second goal of this paper -- a  
proposal for a semiclassical evaluation of the nonzero temperature  
partition function.

%%%%%%%%%%%%%%%%%%%%%%%%%%%%%%%%%%%%%%%%%%%%%%%%%%%%%%%%%%%%%%%%  
\section{A string theory dual to hadronic states}  
%%%%%%%%%%%%%%%%%%%%%%%%%%%%%%%%%%%%%%%%%%%%%%%%%%%%%%%%%%%%%%%%%%%%%  
\label{annulons}  
In the framework of the AdS/CFT some supergravity solutions have been  
constructed that are dual to confining gauge theories. Most notable are  
the Klebanov-Strassler (KS) \cite{ks} and Maldacena-N\'u\~nez (MN) \cite{mn}  
solutions. Generalizing some of the ideas proposed in \cite{bmn}, a  
particular sector of these theories has been isolated. The resulting  
string theory is exactly solvable and describes hadronic excitations  
of the gauge theory \cite{hadrons}. Since the Penrose-G\"uven limit
of either KS or MN backgrounds entails boosting along a certain compact
direction transverse to the gauge theory directions, the hadronic states 
that are dual to the string theory modes necessarily have large U(1) charge.
These hadronic states that are selected upon taking the Penrose-G\"uven limit
may be viewed as ripples on an infinitely heavy 
configuration with a large U(1) charge, called annulon \cite{hadrons}.

In this section we will consider the theory emerging as a limit of the  
Maldacena-N\'u\~nez solution since it is technically simpler to deal  
with, and since it already contains all of the features believed 
to be universal for  
duals of confining theories.  The light-cone  
Hamiltonian in question is \cite{hadrons}:  
\bea  
\label{TheHamiltonian}  
H&=&{P_{i}^2 \over 2p^+ } + {P_{4}^2 \over 2p^+ }+{1\over 2 \a'  
    p^+}\sum\limits_{n=1}^\infty n(N_n^i+N_n^4) \nonumber \\  
&+&{1\over 2\a'p^+}\sum\limits_{n=0}^\infty \left( w_n^a\,N_n^a+ w_n^b\,N_n^b\right)  
    \nonumber \\  
&+& {1\over 2\a'p^+}\sum\limits_{n=0}^\infty \left(\o_n^\a {\cal S}_n^\a+\o_n^\b {\cal S}_n^\b\right).  
\eea  
where $i=1,2,3$, $a=5,6$, $b=7,8$, $\a=1,2,3,4$ and $\b=5,6,7,8$;    
$N_n^s=a_n^{s,\dagger} a_n^s+ \tilde{a}_n^{s,\dagger} \tilde{a}_n^s$      
and ${\cal S}^s=S_n^{s,\dagger} S_n^s+ \tilde{S}_n^{s,\dagger}  
    \tilde{S}_n^s$   
are bosonic and fermionic occupation numbers  
respectively which include both left and right movers.   
The frequencies that appear in the Hamiltonian are  
\bea  
w_n^a &=& \sqrt{n^2+(m_0 p^+ \a')^2}, \qquad w_n^b=\sqrt{n^2+{1\over  
    9}(m_0 p^+ \a')^2}, \nonumber \\  
\o_n^\a &=&\sqrt{n^2+{1\over 9}(m_0 p^+\a')^2}, \qquad \o_n^\b =\sqrt{n^2+{4\over 9}(m_0 p^+\a')^2}.  
\eea  
  
The general structure of the light-cone Hamiltonian is rather  
simple. In ten dimensions the bosonic sector contains   
eight physical degrees of freedom. Generically three of these eight  
are massless. These three fields can be traced to part of the  
Poincare symmetry in the original background. The remaining five  
degrees of freedom represent specific excitations of the ground state  
and they are generically massive bosons in two dimensions. Some of the  
values of the masses can be traced back to symmetries of the original  
background before taking the limit. In other words, the Hamiltonian  
receives 
contributions from the momentum and stringy excitations in the  
spatial directions of the field theory (index  
$i=1,2,3$),$H_{\parallel}$,   
and a contribution from the massive ``zero''  
modes and excitations of the internal directions (index  
$s=4,5,6,7,8$), $H_{\perp}$.  
  
There are  two important features which both the  
bosonic MN and KS Hamiltonians share. First, both theories have  
the same $H_{\parallel}$. Second, they have two worldsheet bosons with  
mass $p^+\alpha' m_0$. Where $m_0$ is defined in such a way that the  
energy $E$ of the string theory vacuum state is $Jm_0$, where $J$ is a  
large number representing an internal $U(1)$ charge in the dual gauge theory. The two  
theories then share the fact that the lowest-lying mode of the two  
massive bosons  
shifts $E$ by exactly $m_0$ (see \cite{hadrons} for a more complete  
description and notation). There is a feature that was not completely  
understood in \cite{hadrons} and that is not believed to be universal:  
one of the transverse excitations is massless.  
  
The expression (\ref{TheHamiltonian}) is, in principle, sufficient to
 calculate the thermal  
partition function. However, it is very  convenient to use a path  
integral approach (see \cite{pv}). Following this approach we use that  
the building blocks of the full partition function are the partition  
function of a massive boson and a massive fermion on the torus  
    described by the modular parameter $\tau$:  
\bea  
z_{lc}^{(0,0)}(\tau, m_0) &=& \left[\prod_{n_1,n_2\in {\bf Z}}    
Im \tau \left({(\frac{2\pi}{4Im \tau})^2 |n_1 \tau - n_2|^2  
    +{m_0^2\,\b^2\over Im \tau^2}    
}\right)\right]^{-1} \\
&=&\exp\left[
-\pi Im \tau\sum_{n\in {\bf Z}} \sqrt{n^2+m^2}\right] 
\bigg[\prod_{n\in {\bf Z}}\left(1-\exp[2\pi (-Im \tau \sqrt{n^2+m^2}+iRe \tau  n)]
\right)\bigg]^{-1}.\label{z00} \nonumber 
\eea
\bea
\label{z012}
z_{lc}^{(0,1/2)}(\tau,m_0)\!\!\!\! &=&\!\!\!\! \prod_{n_1,n_2\in {\bf Z}}    
Im \tau \left({(\frac{2\pi}{4Im \tau})^2 \left|n_1 \tau    
+ \frac{2 n_2+1}{2}\right|^2 +{m_0^2\,\b^2\over Im \tau^2}  }\right) \\
\!\!\!\!&=&\!\!\!\! \exp\left[
\pi Im \tau \sum_{n\in{\bf Z}} \sqrt{n^2+m^2}\right]
\prod_{n\in{\bf Z}} \left(1+\exp[2\pi (-Im \tau \sqrt{n^2+m^2}
+i Re \tau n)]\right). \nonumber 
\eea  
where $Im\tau = \beta/(2\pi p^+)$.
The free energy of a gas of non-interacting strings 
is:  
\be
F=-{1\over \b}{\rm Tr}\bigg[(-1)^{\bf F}\ln(1-(-1)^{\bf F})e^{-\b E}\bigg].
\ee
We evaluate the above expression explicitly by series expanding the $\ln$ in
and grouping the sum over even and odd integers:
\bea    
\label{e}  
{F} &=&-{1\over 2\pi \, l_s}    
\int\limits_0^\infty \frac{d Im \tau}{Im \tau{}^2}\int\limits_{-1/2}^{1/2} 
d Re \tau    \bigg\{
\sum\limits_{r=\rm {odd}}\, \exp(-\frac{\beta^2\,  
  r^2}{2\pi\,\alpha'\,Im \tau}) \nonumber \\  
&\times&\left[Im \tau^{-1/2}|\eta(\tau)|^{-2}\right]^4  
\left[z_{lc}^{(0,0)}(\tau,\frac{m_0\,\beta\, r}{Im \tau})\right]^2  
\left[z_{lc}^{(0,0)}(\tau,\frac{m_0\,\beta\, r}{3 Im \tau})\right]^2  
\nonumber \\  
&\times& \left[z_{lc}^{(0,1/2)}(\tau,\frac{m_0\,\beta\, r}{3Im \tau})
\right]^4  
\left[z_{lc}^{(0,1/2)}(\tau,\frac{2m_0\,\beta\, r}{3 Im \tau})\right]^4
\nonumber\\
&+& {\sum\limits_{r=\rm {even}}}'\,\exp(-\frac{\beta^2\,  r^2}{2\pi\,\alpha'
\,Im \tau})\left[Im \tau^{-1/2}|\eta(\tau)|^{-2}\right]^4  
\left[z_{lc}^{(0,0)}(\tau,\frac{m_0\,\beta\, r}{Im \tau})\right]^2  
\left[z_{lc}^{(0,0)}(\tau,\frac{m_0\,\beta\, r}{3 Im \tau})\right]^2  
\nonumber \\&\times& \left[z_{lc}^{(0,0)}(\tau,\frac{m_0\,\beta\, r}
{3Im \tau})
\right]^{-4}  
\left[z_{lc}^{(0,0)}(\tau,\frac{2m_0\,\beta\, r}{3 Im \tau})\right]^{-4}
\bigg\}   .  
\eea  
A simplifying  way to look at the above expression (\ref{e}) is to realize 
that it   
is nothing but the thermal one loop free energy of a string  theory with the 
following  
content: four  massless bosons, two bosons with mass $m_0$, two  
 bosons with mass $m_0/3$; four fermions with mass $m_0/3$ and four more  
 fermions with mass $2m_0/3$. Note the change in the spin structure of the
 fermionic contribution to the partition function which can be explained
 by the action of $(-1)^{\bf F}$.   
  
From expression (\ref{e}) and using some of the modular properties  
discussed in \cite{pv},  we conclude that this string theory has a Hagedorn  
temperature given by the following equation:  
\be  
-\tilde{T}_s\b_H^2 +{4\over 3}\pi -8\pi \g_0(m_0\b_H)-8\pi  
\g_0({m_0\over 3}\b_H)+ 16\pi \g_{1/2}({m_0\over 3}\b_H)+ 16\pi  
\g_{1/2}({2m_0\over 3}\b_H)=0,  
\ee  
where $\gamma_0 (M)= \sum_{n\in {\bf Z}} \sqrt{n^2+M^2}$ is the Casimir energy
of a massive bosonic degree of freedom, while 
$\gamma_{1/2}(M)  = \sum_{n\in {\bf Z}+1/2} \sqrt{n^2+M^2}$
is the Casimir energy of a fermionic degree of freedom, with spin structure
$(1/2,0)$. In the previous relation we have used the mapping between string 
theory and gauge theory  
quantities obtained in \cite{hadrons}, in particular $1/2\pi  
\a'=T_s/J=\tilde{T}_s$. The generalization of the Penrose-G\"uven  
limit taken in \cite{hadrons} amounts, on the gauge theory side, to  
sending the string tension and the $U(1)$ charge of the state to  
infinity:   
$T_s,J \to \infty$ with the ratio $\tilde{T}_s=T_s/J$ held fixed.

Notice that for small $m_0$ the value of the Hagedorn temperature  
reduces to that of IIB strings in flat space. For any nonvanishing $m_0$ we  
find an increase in the value of the Hagedorn temperature.   
  
More important is the regime of very large $m_0$ since it has proved  
to be very relevant in gauge theory applications of the BMN  
construction. As shown in \cite{pv} for very large values of $m_0$ the  
functions $\g_0$ and $\g_{1/2}$ exponentially vanish and the main  
contribution comes, exclusively, from the directions where the hadrons  
can scatter, that is, the flat directions. The density of states in the
region  ($m_0\to \infty )$ is
thus 
\be
\label{density}
d(E)\approx \exp\left(\sqrt{4\pi\over 3}\,{E\over \tilde{T_s}^{1/2}}\right).
\ee
This is in contrast to the  
situation for the plane wave \cite{pv,therest,gost} where the Hagedorn temperature  
goes to infinity in this limit. It can be shown that the Hagedorn
temperature is a monotonic function of $m_0$ \cite{gost}. Thus, 
as we vary $m_0$ the Hagedorn temperature range is
$\tilde{T_s}^{1/2}/\sqrt{4\pi}\le T_H \le \sqrt{3}
\tilde{T_s}^{1/2}/\sqrt{4\pi}$.

%%%%%%%%%%%%%%%%%%%%%%%%%%%%%%%%%%%%%%%%%%%%%%%%%%%%%%%%%%%%%%%%%%%  
\section{The role of temporal windings}  
%%%%%%%%%%%%%%%%%%%%%%%%%%%%%%%%%%%%%%%%%%%%%%%%%%%%%%%%%%%%%%%%%%%%  

$${}$$

There are various effective ways in string theory to think about the Hagedorn  
temperature. It was realized by Polchinski \cite{polchinski} that the  
standard field theoretic approach of compactifying time to study  
thermal properties of field theory could be extended to string theory  
as well, by simply compactifying the target space time. This opens up  
the possibility of interpreting the Hagedorn density of states as  
caused by the windings in the temporal direction. This development was  
extended in  \cite{mcclain} and further  
elaborated upon in \cite{aw}.  
  
Let us first, for completeness, recall the structure of the partition  
function for a compactified boson on the torus \cite{cft}.   
We choose the metric with torus topology to be   
\be  
ds^2=|d\s_1+\tau d\s_2|^2=d\s_1^2+|\tau|^2d\s_2^2 +2({\rm Re}\, \tau) d\s_1d\s_2.  
\ee  
The worldsheet action for a bosonic field is  
\be  
S={1\over 4\pi \a'}\int d\s_1 d\s_2 \sqrt{\g}\,\g^{\a\b}  
\partial_\a X \partial_\b X.  
\ee  
We are interested in considering configurations with nonzero winding  
number and therefore consider    
\bea  
X&=&X^{classical}_{m,n}+X^{quantum}, \nonumber \\  
X^{classical}_{m,n}&=& m\b\,\s_1 + n\b\, \s_2, \nonumber \\  
X^{quantum}&=&\sum_{n_1, n_2}X_{n_1,n_2}e^{2\pi i (n_1\, \s_1+n_2\, \s_2)} .  
\eea  
Note that $X^{classical}$ satisfies the equation of motion  
$\pd_\a(\sqrt{\g}\g^{\a\b}\pd_b X)=0$, and  
determines a topological sector $(m,n)$; hence the notation. The  
quantum part is single-valued and corresponds precisely to the  
expansion of a bosonic field on the torus.   
Evaluating the action on the classical configuration we have  
\be  
S[X^{classical}_{m,n}]=\b^2\,{|m\tau-n|^2\over 4\pi\,\a' Im \tau}.\nonumber   
\ee  
In calculating the partition function, the part coming from the  
single-valued $X^{quantum}$ is modular invariant by itself. To render  
the full partition function modular invariant we need to sum over all  
pairs $(m,n)\in {\bf Z}^2$. The free energy written in a manifestly modular
invariant way is\footnote{Including the sum over $(m,n)\in {\bf Z}^2$
  for modular invariance 
  amounts effectively to account for  multi-string states which is
  nothing but taking the logarithm  of the 
  partition function of a noninteracting string gas.}

\bea  
\label{free}
{F} &=&-\frac 1\beta Z_{T^2}\nonumber\\
&=&-{1\over 2\pi \, l_s}  
{\sum\limits_{m,n \in {\bf Z}}}'
\int_{\cal F} \frac{d^2 \tau}{Im\tau^2} 
\,\,{1\over  
  Im \tau^{1/2} |\eta(\tau)|^2}{\sum\limits_{m,n \in {\bf Z}}}'
\exp\left(-{\b^2\over  
  4\pi\,\a' Im \tau}|m\tau-n|^2\right).  
\eea  
Note that in this case the free energy factorizes as  
\be  
{F} =  {F}_{quantum}\,\,{F}_{classical}.   
\ee  
This factorization turns out to be a property of flat space but the  
idea of computing the full partition function by first finding a  
classical configuration which incorporates the temporal windings, 
and then considering quantum fluctuations  
around it will be central to our proposal for computing the thermal  
partition function semiclassically.   
  
At this point we are ready to revisit the partition function of the  
annulons in the previous section. Our intention is to cast the result  
of section \ref{annulons} in a way that allows an interpretation as the  
partition function calculated based on a classical solution and  
quantum fluctuations around it. There are, however, a few subtleties 
to take into account. The natural candidate for a classical  
solution would be the temporal coordinate playing the role of a  
compactified boson described above. However, in the light-cone  
treatment of section \ref{annulons} the time target space coordinate  
is gauged away. A fully covariant approach requires dealing with the RR  
3-form field. We are content though with a partially covariant approach,
in the sense that the bosons are treated covariantly, while the fermions
are $\kappa$ gauge fixed.  
  
To make explicit the emergence of a semiclassical solitonic configuration,  
it is crucial to note that  
although we have suggestively written the integration  
variables as $Re \tau$ and $Im \tau$, the  partition function (\ref{e}) 
has no  
obvious modular properties since we are integrating over the strip    
\be  
\label{s}  
E: \quad Im \tau>0, \quad -{1\over 2}< Re \tau  < {1\over 2}.  
\ee  
An interesting result obtained for strings in flat space \cite{tan} but  
that can be generalized to the current situation is that the torus partition  
function (\ref{e} and \ref{free}) can be written in an explicitly  
modular invariant way:  
\bea  
\label{f}  
Z_{T^2}&=&  
{\beta\over 2\pi \, l_s}    
\int\limits_{\cal F}\frac{d Im \tau}{Im \tau{}^2}\int d Re \tau   
\sum\limits_{m,n}{}' \exp\left(-{\b^2|m\tau  
  -n|^2\over 4\pi\,\a'Im \tau}\right)\nonumber \\  
&\times&\left[Im \tau^{-1/2}|\eta(\tau)|^{-2}\right]^4  
\left[z_{lc}^{(0,0)}(\tau,\frac{m_0\,\beta\,}{Im \tau}|m\tau  -n|)\right]^2  
\left[z_{lc}^{(0,0)}(\tau,\frac{m_0\,\beta\, }{3 Im \tau}|m\tau  -n|)
\right]^2  
\nonumber \\  
&\times& \left[z_{lc}^{(b1,b2)}(\tau,\frac{m_0\,\beta\, }
{3Im \tau}|m\tau  -n|)\right]^4  
\left[z_{lc}^{(b_1,b_2)}(\tau,\frac{2 m_0\,\beta\, }{3 Im \tau}|m\tau  -n|)
\right]^4  
\eea  
where $m$ and $n$ are integers and we exclude $m=n=0$; also $b_1=(1- (-1)^m)/4,
 b_2= (1- (-1)^n)/4$ denote the fermion spin structure in a given topological
sector.
 Notice that the 
integration is  over the fundamental domain  
\be  
\label{fd}  
{\cal F}: \quad |\tau| > 1, \quad -{1\over 2}< Re \tau  < {1\over 2}.  
\ee  
The proof of the above presentation of the partition function relies  
on the definition of fundamental domain: a region of the upper  
half-plane such that no two points  are related by a modular  
transformation and any point outside of it can be reached by a modular  
transformation. Using this definition one is to write the strip in  
terms of the fundamental domain ${\cal F}$.   
  
It is the exponent in the first line above (\ref{f}) that can be  
interpreted as the classical action of temporal winding modes.
The most important result of this manipulation is the  
explicit possibility of interpreting the partition function of a  
string theory in a curved background with RR form field as obtained  
from quantum fluctuations around a classical solution which involve  
windings of the spacetime temporal direction.    
  
One interesting observation is that we see that the total partition  
function no longer factorizes as the product of $Z_{quantum}$ and  
$Z_{classical}$, confirming that this factorization was indeed an artifact of  
flat space. Nevertheless, the dependence of $Z_{quantum}$ on $(m,n)$ is  
rather simple, it reduces to:     
  
\be  
m_0\to m_0|m\tau-n|.  
\ee  
   
%%%%%%%%%%%%%%%%%%%%%%%%%%%%%%%%%%%%%%%%%%%%%%%%%%%%%%%%%%%%%%%%%%%%%%%%%%%%  
\section{A semiclassical evaluation of $Z_1$}  
%%%%%%%%%%%%%%%%%%%%%%%%%%%%%%%%%%%%%%%%%%%%%%%%%%%%%%%%%%%%%%%%%%%%%%%%%%%%%% 
$${}$$
 
Let us assume that we have a supergravity background dual in the
AdS/CFT sense to a gauge theory. The full string theory in such
backgrounds is not known. However, for a semiclassical treatment the
sigma model action is needed only up to quadratic   terms and it is
given by (we follow \cite{tr}):
\bea  
S&=&{1\over 4\pi \a'}\int d\s_1 d\s_2 \sqrt{\g}\,\bigg[(g_{\m\n}\g^{\a\b}+b_{\m\n}\epsilon^{\a\b})  
\partial_\a X^\m \partial_\b X^\n \nonumber \\  
&+&i (\g^{\a\b} \delta_{IJ} -   \epsilon^{\a\b}  
(\rho_{3})_{IJ}) \partial_\a    X^{m } \bar \theta^I \Gamma_{m} D_\b  
\theta^J  \bigg],    
\eea  
where  $\theta^I$ ($I$=1,2)  are the  two  
real positive chirality  10-d  MW spinors  and  
$D_b$ is the pullback to the world-sheet of the supergravity  
covariant derivative in the variation of  
the gravitino :  
\begin{equation}     
D_\a =  \partial_\a   + { 1 \over 4}  
\partial_\a X^m \big[\     (\omega_{ \mu \nu m}   - {1\over 2}  H_{ \mu  
\nu m} \rho_3) \Gamma^{ \mu  \nu}  +  ( { 1 \over  3!}  
F_{\mu\nu\lambda} \Gamma^{\mu\nu\lambda} \rho_1  +   { 1 \over 2\cdot  
5!} F_{\mu\nu\lambda\rho\kappa}   \Gamma^{  
\mu\nu\lambda\rho\kappa}\rho_0 ) \Gamma_m  \  \big]    
\end{equation}  
where  the $\r_s$-matrices in the $I,J$ space  are the Pauli matrices  
$ \rho_1 = \sigma_1$, $\rho_0 = i \sigma_2$, $\rho_3 =\sigma_3$ .    
  
The basic idea for a semiclassical estimation of the nonzero  
temperature partition function is schematically as follows. We  
consider the existence of winding temporal modes a crucial  
ingredient and thus, include them as part of the ansatz. In  
general, due to the nontriviality of the warp factor $(g_{00})$, 
this ansatz will fail and the  
``minimal'' modification one is to consider as the classical solution is   
\be  
X^0=m\b \s_1+ n\b \s_2,  \qquad r=r(\s_1,\s_2).  
\ee  
Assuming that the background contains only a nontrivial metric and RR  
3-form (for simplicity we are keeping in mind the solution of   
\cite{mn}), the nontrivial equations of motions are:  
\bea  
\label{sys}  
\partial_\a(\sqrt{\g}\g^{\a\b}g_{00}\partial_\b X^0)&=&0, \nonumber \\  
\partial_\a\left(\sqrt{\g}\g^{\a\b} g_{rr}\partial_\b r\right)-{1\over  
  2}\partial_rg_{00}\bigg[\sqrt{\g}\g^{\a\b}\partial_\a X^0\partial_\b X^0\bigg]&=&0.  
\eea  
Given a classical solution $(X^0, r)$, the  
action can be evaluated as:  
\bea  
S[X^{classical}_{m,n}]&=& {1\over 4\pi \a'} \int d\s_1 d\s_2 {1\over  
  Im \tau}\bigg[\b^2(n^2+m^2|\tau|^2-2({\rm Re}\, \tau)\,  
  m\, n)g_{00} \nonumber \\  
&+&  g_{rr} \, (|\tau|^2\,\dot r^2+r'^2-2 ({\rm Re})\, \dot r  
  r')\bigg].  
\eea  
The solution to the coupled system of differential equations
(\ref{sys}) is in  
general very involved. There is, however, a case were a simple  
solution exists. Let us assume the existence of a point $r_0$ such  
that :  
\be  
\label{cond}  
\exists\, r_0: \,\, g_{00}(r_0)\ne 0,\quad   
\partial_r g_{00}(r_0)=0, \quad \partial_r g_{rr}(r_0)=0.   
\ee  
Then, one solution to the system (\ref{sys}) is simply  
\be  
\label{sol}  
X^0=m\b \s_1+ n\b \s_2,  \qquad r(\s_1,\s_2)=r_0.  
\ee  
Now the question is -- what kind of supergravity backgrounds in the  
context of the AdS/CFT admit such behavior? In a series of papers  
Sonnenschein and collaborators \cite{cobiwl} have shown that the conditions  
(\ref{cond}) are precisely the conditions the supergravity background  
must satisfy in order for the dual gauge theory to be  
confining. This analysis is based on the AdS/CFT  evaluation of the
expectation value  of the Wilson loop introduced in \cite{wlads}. Moreover, the
tension in the gauge theory dual is given by  
$T_s=g_{00}(r_0)/2\pi \alpha'$. Thus,  
a solution of the type (\ref{sol}) exists for any supergravity  
background dual to a confining theory. The classical action then becomes  
\be  
S[X^0_{classical}, r=r_0]=T_s{\b^2 |m\tau-n|^2\over 2 Im \tau}.  
\ee  
The most salient feature of this classical action is that the effective  
string tension that appears is precisely that of the confining gauge  
theory.   
  
%%%%%%%%%%%%%%%%%%%%%%%%%%%%%%%%%%%%%%%%%%%%%%%%%%%%%%%%%%%%%%%  
\subsection{Fluctuations}  
%%%%%%%%%%%%%%%%%%%%%%%%%%%%%%%%%%%%%%%%%%%%%%%%%%%%%%%%%%%%%%%%  
$${}$$ 
  
Let us now turn to the evaluation of $Z_{quantum}$ by  
considering quadratic fluctuations around the classical solution  
(\ref{sol}). For simplicity and concreteness we will consider the  
background describing  $N$ D5-branes wrapped on $S^2$ \cite{mn,cv} but
the method applies to any supergravity dual to a confining gauge theory. 

The metric of the solution is 
\bea  
ds^2&=&e^\p\bigg[dx^adx_a+\a'g_s  
N(d\tau^2+e^{2g(\tau)}(e_1^2+e_2^2)+{1\over 4} (e_3^2+e_4^2+e_5^2)\bigg],  
\nonumber \\  
e^{2\p}&=&e^{-2\p_0}{\sinh 2\tau\over 2e^{g(\tau)}}, \nonumber \\  
e^{2\,g(\tau)}&=&\tau\coth 2\tau -{\tau^2\over \sinh^2 \, 2\tau}-{1\over  
  4}, \nonumber \\  
\eea  
where,   
\bea  
e_1&=&d\theta_1, \qquad e_2=\sin\te_1 d\p_1, \nonumber \\  
e_3&=&\cos\psi\, d\te_2+\sin\psi\sin\te_2\, d\p_2 -a(\tau) d\te_1,  
\nonumber \\  
e_4&=&-\sin\psi\, d\te_2+\cos\psi\sin\te_2\, d\p_2 -a(\tau) \sin \te_1d\p_1,  
\nonumber \\  
e_5&=&d\psi +\cos\te_2\, d\p_2 -\cos\te_1d\p_1, \quad  
a(\tau)={\tau^2\over \sinh^2\tau}.  
\eea   
The position referred to as $r_0$ in the previous subsection is
$\tau=0$. Therefore, we will expand the metric around that value. A  
fairly nontrivial fact we use is that $e_3^2+e_4^2+e_5^2$ evaluated at
$\tau=0$ is the  
round $S^3$ with radius $1/\sqrt{2}$. This can be verified  by  
writing this line element as the $SU(2)$ invariant line element via a direct 
map  
involving a specific $SU(2)$ matrix \cite{mt}. 
Therefore, we parametrize this
round $S^3(\te,\phi,\psi)$ by its Euler angles. 
Next, choosing to expand near the classical
value $\te=\pi/2$ leads us effectively to ${\bf
  R}^3(y^1,y^2,y^3)$. Near $\tau=0$ we have that $e^{2g}\approx
\tau^2$ and therefore we combine the $\tau$-direction with
$S^2(\te_1,\p_1)$ into ${\bf R}^3(\tau^1,\tau^2,\tau^3)$ in Cartesian
coordinates. The end result for the bosonic quadratic action  is
\bea  
S_{2b}&=&S[X^0_{classical}, r=r_0]+{1\over 2\pi \a'}\int d\s_1 d\s_2  
  \sqrt{\g}\g^{\a\b}\left(\pd_\a X^a \pd_\b X^a g_{00}  \right.   
\nonumber \\  &+&\left.
\a' g_s\, Ng_{00}\big[\pd_\a\tau^i \pd_\b \tau_i+{1\over 4}\pd_\a y^i  
\pd_\b y_i\big] 
+{4 \beta^2\over 9 Im\tau^2}g_{00} |m\tau-n|^2 \tau^i\tau_i\right)  
\eea  
where $a=1,\ldots,4$ and $i=1,2,3$. 

To read off the mass term generated in the $\tau_i$ directions, we must
first rescale the fields $\tau_i$ such that the kinetic term is canonically 
normalized.
The bosonic part of the action yields seven massless  
fields, three massive fields with mass $(2/3) \beta
\sqrt{\frac{1}
{\a' g_s N}}|m\tau-n|/Im\tau$, 
and two diffeomorphism ghosts.
When evaluating the one loop partition function, the ghost contribution
will cancel out the contribution of two of the massless fluctuations, leaving
us with three massive and four massless bosonic physical degrees of freedom.

Let us turn now to the fermionic degrees of freedom.
The part quadratic in fermions can be expanded in the presence of a RR
3-form field strength and around (\ref{sol}) to give:
\bea
S_{2f}=\frac i{2\pi\alpha'}
\int \bar\theta^I (\sqrt{\g}\g^{\a\b}\delta^{IJ}-\epsilon^{\a\b}
\sigma_3^{IJ}) \partial_\a X^0 \Gamma_{\underline 0} 
e^{\underline 0}_0 (\delta^{JK}\partial_\b +\frac{1}{8 \cdot 3!}
e^{\phi}\sigma_1^{JK} \Gamma^{\mu_1\mu_2\mu_3}F_{\mu_1\mu_2\mu_3}
\partial_\b X^0\Gamma_{\underline 0}e^{\underline 0}_0)\theta^K\nonumber\\
\eea
where the 3-form is given by
$F_{(3)}=-{1\over 4}g_s N dy_1\wedge dy_2 \wedge dy_3$
and we denoted the vielbeins by $e_{\mu}^{\underline m}$: 
$e_{\mu}^{\underline m} e_{\nu}^{\underline n}
\eta_{\underline m\underline n}=g_{\mu\nu}$.

This action is still invariant under $\kappa$ symmetry. To truncate to the
physical degrees of freedom, we choose the $\kappa$
gauge 
\be
\Gamma^+ \theta^I=0.
\ee
Following the analysis of \cite{dgt}  
one is left then with a theory which is free of
conformal anomaly\footnote{
It was noted in \cite{dgt} that the conformal anomaly vanishes for the
GS string in the $\kappa$ gauge in a flat target space background, 
where the analysis reduces to considering 
the central charge contribution of the
ten scalars, one pair of ghosts and eight pairs of GS fermions:  
$10-26+8\times 4\times
\frac 12=0$. The contribution to the conformal anomaly of a GS 
fermion pair
is 4 times that of a 2d Majorana worldsheet spinor \cite{dgt}.
More generally, in curved backgrounds, 
the vanishing of the conformal anomaly is proven
by computing the effective Liouville action.}.

Then the $\kappa$ gauge fixed action $S_{2f}$
becomes
\bea
S_{2f}=\frac i{2\pi\alpha'}
\int &&\bar\theta^I (\sqrt{\g}\g^{\a\b}\delta^{IJ}-\epsilon^{\a\b}
\sigma_3^{IJ})
\partial_\alpha X^+ \Gamma^- (g_{00})^{-1/2}(\delta^{JK}\partial_\b
\nonumber\\
&+&
\frac{1}{8 \cdot 3!}
e^{\phi}\sigma_1^{JK} \Gamma^{\underline i_1' \underline i_2' 
\underline i_3'}F_{i_1' i_2' i_3'}\partial_\b X^+ \Gamma^- (g_{00})^{-1/2})
\theta^K
\eea
where we have used that the classical solution is characterized by 
non-vanishing $X^\pm$ as well as the gauge condition.
Next notice that the term proportional to $F_{(3)}$ cancels by the gauge
condition. We are left therefore with 8 massless GS fermions.

One last ingredient needed for the finite temperature partition function
relates to the fact that in the path integral, the boundary 
conditions obeyed by the thermalized 
fermionic degrees of freedom in a given soliton sector, characterized by 
$(m ,n )$ winding numbers, are \cite{carlip}
\bea
\theta(\sigma_1+1,\sigma_2)&=&(-1)^{m}\theta(\sigma_1,\sigma_1)\nonumber\\
\theta(\sigma_1,\sigma_2+1)&=&(-1)^{n}\theta(\sigma_1,\sigma_1).
\eea
Finally, putting  all the pieces together, the one loop  finite 
temperature partition function will be given by
\bea
Z_{T^2}\!\!\!\!\!&=&\!\!\!\!\!\!\!{\sum_{m,n\in {\bf Z}}}'\frac{\beta}{2\pi l_s}
\int_{\cal F} 
d^2 \tau \frac 1{Im \tau^2} e^{-\frac{\beta^2 g_{00}}{4\pi\alpha'}
\frac{|m\tau-n|^2}{Im \tau}}
z_{0,0}^b(\tau, 0)^5 z_{0,0}^b(\tau, M^2=
\frac 49 \beta^2 \frac{|m\tau -n|^2}{Im\tau^2} 
\frac{1}{\a' g_s N})^3 z_{b_1, b_2}^f(\tau,0)^8\nonumber\\
\label{part}\eea
where  \cite{pv}
\be
z^b_{0,0}(\tau , M)=
e^{-\pi Im\tau \sum_{l\in {\bf Z}} \sqrt{l^2 +M^2}}{\prod_{l\in {\bf Z}}}
\bigg(1-e^{-2\pi Im\tau \sqrt{ l^2 +M^2}+2\pi i Re\tau l}\bigg)^{-1}
\ee
denotes the contribution of a bosonic degree of freedom with mass $M$
and regular boundary conditions, while
\bea 
z_{b_1, b_2}^f(\tau,M)&=&e^{\pi Im\tau \sum_{l\in {\bf Z}}\sqrt{(l+ b_1)^2+M^2}}
{\prod_{l\in{\bf Z}}} (1-e^{-2\pi
Im\tau \sqrt{ (l+b_1)^2 +M^2}+2\pi i Re\tau (l+b_1) -
2\pi i b_2}),
\nonumber\\
\eea
denotes the contribution of a GS fermion, with mass $M$ and in the soliton
sector $m,n$, with twisted boundary conditions 
$b_1=(1-(-1)^m)/2$ and $b_2 = (1- (-1)^n)/2$.

The partition function (\ref{part}) has a potential divergence as $Im\tau
\rightarrow \infty$. Reading off only the dominant exponentials in this limit
we find that for $m=1$ the integrand of the partition function 
\be
Z_{T^2}\approx \int 
e^{-\frac{\beta^2 g_{00}}{4\pi\a'} Im \tau} e^{-\pi
Im\tau\sum_{l\in {\bf Z}} (5 l +3\sqrt {l^2 +
\frac 4{9} \beta^2 \frac{1}{\a' g_s N}} - 8 (l+\frac 12))},
\ee
becomes divergent at a critical temperature $T_H$: 
\be
\frac{1}{4\pi\alpha'}\beta_{H}^2 g_{00}=-2\pi
\bigg(5 \gamma_0 (0) +3 \gamma_0 (2\beta_H \sqrt{\frac{1}{\a' g_s N}}/3)
- 8 \gamma_{1/2}(0)\bigg).
\ee
The Hagedorn temperature is none other than the temperature where the 
first winding soliton becomes tachyonic \cite{aw}.

In the limit of very large $1/(\a' g_s N)$ we find that the density of 
states is
given by
\be
\label{hag}
d(E)\approx \exp\left(\sqrt{3\pi}\,\, {E\over T_s^{1/2}}\right).
\ee
Note that the density of states depends, as advertised,  on the gauge 
theory quark-antiquark string tension.

%%%%%%%%%%%%%%%%%%%%%%%%%%%%%%%%%%%%%%%%%%%%%%%%%%%%%%%%%%%%%%%  
\section{Conclusions}  
%%%%%%%%%%%%%%%%%%%%%%%%%%%%%%%%%%%%%%%%%%%%%%%%%%%%%%%%%%%%%%%%  
$${}$$

In our proposal there is certainly an ambiguity in the choice of the classical
solution around which to expand. We have argued for the need for
temporal winding modes but this does not exclude other
configurations. In fact, in principle the partition function evaluated
semiclassically should include a sum over all classical solutions. 
Effectively, in this paper we have considered two  
such expansions. The annulons are precisely an expansion around a  
classical solution which corresponds to large $U(1)$ flavor  
charge in the field theory. The advantage of the simple choice advocated 
in the second  
part of the paper is precisely the absence of any charge, since   
in order to hope to collect information close to the pure  
${\cal N}=1$ sector which is neutral under any charge contained 
in the supergravity  
background we should not involve any charge.  
The only possible charge -- $U(1)_R$ -- is broken by gaugino  
condensate in the IR. Thus, our choice of the classical solution is
justified for the purpose of extracting information about pure ${\cal
  N}=1$ SYM.   
  
Let us comment on the nature of the density of states discussed
here. The ``confinement/deconfinement'' of ${\cal N}=4$ SYM  
on $S^3$ is kinematical, reflecting just the $N^0$ {\it versus} $N^2$  
transition in the free energy \cite{ew}. Similarly, properties 
like a Hagedorn temperature in these conformal theories
\cite{pv,therest,gost,nothing,bp} should be understood as a kinematical effect. 
The confining properties we have discussed  
here correspond, on the contrary, to theories in the confining phase with dynamical  
confinement. Namely, we are studying string theory backgrounds  
believed to be dual to embeddings of ${\cal N}=1$ SYM into string theory  
in such a way that the dynamical confining properties remain. Thus,
our calculation are expected to be more than a kinematical effect implied
by the strict $N\to \infty$ limit.   
  
In this paper we have derived the density of states of a string theory  
dual of hadronic states. Since the string theory  is exactly  
soluble we have been able to extract this quantity exactly. The physical  
input needed to obtained the solvable limit (large $U(1)$ charge) shows itself in the final  
answer for the density of states (\ref{density}) by modifying the effective tension to be $\tilde{T}_s=T_s/J$. 

In the second part of the paper, motivated by the exact calculations  
performed in the first part, we proposed a semiclassical evaluation of  
the finite temperature partition function. Our proposal is  
particularly useful for supergravity backgrounds dual to confining  
theories. We carried this semiclassical calculation for the  background  
of $N$ D5 wrapped on $S^2$ whose low energy sector contains pure ${\cal  
  N}=1$ SYM \cite{mn}.  
We showed that generically it gives a Hagedorn density of  
states with the coefficient completely determined by the gauge theory  
string tension  $T_s$ (\ref{hag}). Interestingly, this is precisely the tension $T_s$ of the  
quark-antiquark potential as calculated in the AdS/CFT framework.    

One possible direction is to apply our proposal to other confining
theories, most naturally to the embedding of ${\cal N}=1$ discussed in
\cite{ks}. It would also be interesting to study confining theories 
whose supergravity duals are related to manifolds of $G_2$
holonomy. Finally, for the case of nonconfining theories our proposal
should lead to string theoretic corrections to the Bekenstein-Hawking
entropy.  
  
%%%%%%%%%%%%%%%%%%%%%%%%%%%%%%%%%%%%%%%%%%%%%%%%%%%%%%%%%%%%%%%  
\section*{Acknowledgments}  
%%%%%%%%%%%%%%%%%%%%%%%%%%%%%%%%%%%%%%%%%%%%%%%%%%%%%%%%%%%%%%%%  
$${}$$

We would like to thank   R. Brower, A. Buchel, J. Gomis, C. N\`u\~nez, 
E. Rabinovici,
G. Semenoff, J. Sonnenschein, M. Strassler and A. Tseytlin. 
We are especially thankful to D. Kutasov for raising the question of a
semiclassical estimate for the Hagedorn density of states in confining
theories and to I. Klebanov for a very illuminating discussion.
D.V. would like to thank the  organizers of the QCD and Strings 
workshop and the 
Michigan Center for Theoretical Physics for hospitality during 
the initial stages of this work. 

The research of L.P.Z. is partially 
supported by the U.S. Department of Energy, while D.V. is supported by
DOE grant DE-FG02-91ER40671.
  
%%%%%%%%%%%%%%%%%%%%%%%%%%%%%%%%%%%%%%%%%%%%%%%%%%%%%%%%%%%%%%%  

\end{document}